\documentclass[aps,prl,showpacs,twocolumn]{revtex4}
\usepackage[utf8]{inputenc}
\usepackage{amsmath}
\usepackage{amssymb}
\usepackage{amsthm}
\usepackage{graphicx}
\usepackage{graphics}
\begin{document}

\title{Electronic oscillations in paired polyacetylene chains}

\author{C. R. Muniz$^{1,2}$ and R. N. Costa Filho$^{1,3}$}

\affiliation{$^1$Departamento de F\'{\i}sica, Universidade Federal do Cear\'{a}, Caixa
Postal 6030, Campus do Pici, 60455-760, Fortaleza, Cear\'{a}, Brazil.\\$^2$Universidade
Estadual do Cear\'a, Faculdade de Educa\c c\~ao, Ci\^encias e Letras de Iguatu, Rua
Deocleciano Lima Verde, s/n Iguatu,Cear\'a, Brazil.\\$^3$Department of Physics and
Astronomy,University of Western Ontario London, Ontario, Canada N6A 3K7.}

\date{\today}

\begin{abstract}

An interacting pair of polyacetylene chains are initially modeled as a couple of undimerized polymers described by a Hamiltonian based on the tight-binding model representing the electronic behavior along the linear chain, plus a Dirac's potential double well representing the interaction between the chains. A theoretical field formalism is employed, and we find that the system exhibits a gap in its energy band due to the presence of a mass-matrix term in the Dirac’s Lagrangian that describes the system. The Peierls instability is introduced in the chains by coupling a scalar field to the fermions of the theory via spontaneous symmetry breaking, to obtain a kink-like soliton, which separates two vacuum regions, i.e., two spacial configurations (enantiomers) of the each molecule. Since that mass-matrix and the pseudo-spin operator do not commute in the same quantum representation, we demonstrate that there is a particle oscillation phenomenon with a periodicity equivalent to the Bloch oscillations.

\vspace{0.75cm}
\end{abstract}


\maketitle

Quantum mechanical particles moving under the influence of a
constant electric field and submitted to a periodic potential
oscillate instead of moving with uniform acceleration. This
phenomenon is called Bloch oscillations and were predicted
theoretically by F. Bloch in $1928$ \cite{bloch}. Such
oscillations have never been observed in a natural lattice because 
the characteristic times of the electrons scattering by the lattice
defects, or impurities, are much shorter than the Bloch period. As a
consequence, during a long time these oscillations were seen like
a mere theoretical curiosity to demonstrate the strange properties
of matter, according to quantum mechanics principles. However,
they have been recently observed experimentally in semiconductor
superlattices\cite{Esaki,Helm,Bouchard,Dekorsy,Dekorsy1} and with
larger periods, of the order of ten seconds, in strontium atoms
trapped by laser beams, cooled at temperatures close to absolute
zero. In the former case, the dependence of the Bloch frequency on
the electric field makes the oscillations tunable, yielding a
potential source of coherent high frequency radiation and, in
latter case, the constant external force was the terrestrial
gravitational field itself \cite{Ferrari}.

These oscillations can play an effective and important role in
quantum electronics, due to development of the physics of quasi-unidimensional molecular structures,  as
polyacetylene\cite{roth} and graphene ribbons
\cite{castroneto,Nilsson}.  Polyacetylene consists of a linear chain of carbon atoms, coupled to each other with
alternating simple and double chemical bonds, that can be obtained
via acetylene polymerization. It is an organic polymer with
special electronic properties. Thin films of this polymer produced
under special doping conditions are excellent conductors that can
be used to develop electronic devices in the nanometer
scale\cite{chiang:78:shc}. Due to its dimensionality we can model
the undimerized polyacetylene as a linear chain of carbon atoms
with periodic boundary conditions and a Hamiltonian based on the
tight binding model. In this approach, the valence electrons are
strongly bound to the carbon nucleus, but it has a nonzero
probability of traveling along the chain due to translational
symmetry of the model. As a result the electrons have energy
$E=\pm pv_f$, when calculated close to the Fermi level, indicating
that carriers can be considered as relativistic particles with
zero mass, where the Fermi velocity plays the role of the light
velocity\cite{Zee,Jackiw}. Therefore, that system can be described
by a Dirac Lagrangian without the mass term, where the bi-spinor
entries are the wave functions of electrons propagating to left or
to right in the chain. The zero mass term implies a lack of gap
between the valence and conduction band in the system, generating
a Dirac semimetal. However, the polyacetylene is an insulator or
semiconductor, depending on the density of impurities. Such behavior can
be explained by Peierls instability\cite{su,Peierls}. There are
many technological advantages for polyacetilene to be a
semiconductor instead of a semimetal. The presence of a gap would
increase the on-off ratio for current flow that is needed for many
electronic applications. In graphene, for example, gaps can be
produced by geometrically confining graphene into nanoribbons with
impurities\cite{rai,Han}.

Peierls\cite{Peierls} showed that the undimerized polyacetylene
chain is not stable, at least at ordinary temperatures, and has
introduced a deformation that approximates (\emph{circa} $0.04$
\AA) the carbon ions having double bonds and separates them from
those with single bonds. Such a process of dimerization is called
Peierls instability \cite{su,Peierls}, and generates a gap in the
band structure. In field language, that corresponds to add to the
above mentioned Lagrangian a term which couples a scalar field to the
fermions of the theory. This dimerization process breaks chiral
symmetry spontaneously, transforming such coupling term in a mass
term, generating the one-dimensional soliton structure denominated
\emph{kink}, a topological defect which separates two regions with
stable and different spacial configurations (i.e., vacuum states)
of the molecule, called enantiomers.

Here we propose that bringing two undimerized polyacetilene chains
close together generates a gap in the chain's energy band.
Moreover, the presence of the adjacent chain is responsible for an
electronic oscillation along the other chain. To model this system
we start with two undimerized polyacetylene chains, separated by a
distance $b$ (see Figure 1) along the $y$ direction and composed
of carbon atoms separated by a distance $a$ along the $x$
direction where the electron can propagate throughout the chain.
The Hamiltonian describing the system is given by
\begin{equation} \label{eq:1}
\hat{H}=-t\sum_{i}C_{i}^{\dag}C_{i+1}+\sum_{i}\mathcal{U}(y)C_{i}^{\dag}C_{i}+h.c,
\end{equation}
where $t$ is the hopping integral between sites in the same chain.
The second term in the hamiltonian is the energy of the electrons
due to the presence of the neighboring chain. Here we assume that the electron is
trapped by the paired carbon atoms between the two chains, along
the $y$ direction. Such potential can be represented by a Dirac's double
well potential, where each well is related to the carbon atom, and can be expressed by
\begin{equation}
\label{eq:2}
\mathcal{U}(y)=\frac{-\alpha\hbar^{2}}{2m_{e}b}\left[\delta\left(y-\frac{b}{2}\right)+\delta\left(y+\frac{b}{2}\right)\right].
\end{equation}
The potential depends on the distance $b$ between the two chains.
A particle in such potential has the energy given by
$-\varepsilon_{n}$, with $\varepsilon_{n}>0$, which are the
eigenvalues of the second term in the above Hamiltonian. It is possible to show numerically that such energy eigenvalues are always lesser than the energy of the electron in a single
Dirac's potential well, indicating that,  in normal conditions,
certain diatomic molecules are more stables than the correspondent
single atom ($H_2$ and $H$, for example). The index $n$ of
$\varepsilon_{n}$ is to designate the fundamental and excited
states of the confined electron, if the latter state exists. In
$H_2$ again, there are only the fundamental state \cite{Townsend}.

The expected energy value for this system is

\begin{equation} \label{eq:3}
E(k_x)=-2t\cos{(k_xa)}-\varepsilon_{n}.
\end{equation}

The first term in the above expression can be seen as a particular case of
the one found in Ref.16 when $\Delta=0$. Expanding the above energy around the
$x-$component of the Fermi vector, $k_x=(\pi/2a)$

\begin{figure}
\includegraphics[width=1.0\linewidth]{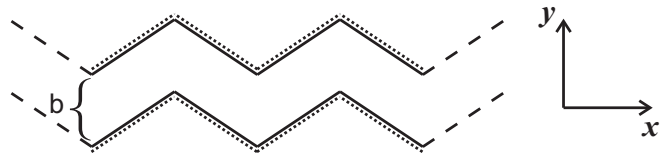}
\caption{The coupled undimerized polyacetylene chains separated by a distance $b$}
\label{fig:1}
\end{figure}


\begin{equation}
\label{eq:4}  {E}=\pm v_{Fx}\overline{p}_x-\varepsilon_{n},
\end{equation}
where the momentum
$\overline{p}_x=p_x-p_{Fx}=\hbar(k_x-k_{Fx})$ is taken relative to
the Fermi level. The Fermi velocity

\begin{equation}
\label{eq:5} v_F=2ta/\hbar,
\end{equation}
and the $\pm$ sign indicates that the electrons can travel in both
directions in the $x-$axis. The electronic energies are quantized
and are different depending on the propagation direction. Two wave
function equations can be constructed related to the energies
in Eq. \ref{eq:4}

\begin{equation}
\label{eq:6} i\hbar\frac{\partial \psi_{R,L}}{\partial t}\pm i\hbar
v_{Fx}\frac{\partial \psi_{R,L}}{\partial
x}\psi_{R,L}+\varepsilon_{n}=0.
\end{equation}
The upper(minus) sign is related to the electron propagation in the
right(left) direction. We can now write the Lagrangian
satisfying the Euler-Lagrange equations which are the wave equations
defined in Eq.\ref{eq:6}

\begin{equation}
\label{eq:7} \mathcal{L}=i\hbar
v_{Fx}\overline\Psi\gamma^{\mu}\partial_{\mu}\Psi-v_{Fx}^2\overline\Psi
M\Psi,
\end{equation}
where $\Psi=\left(\begin{array}{r}\psi_L\\\psi_R\end{array}\right)$
and $\Psi^{\dag}=\left(\psi_L^{\dag},\psi_R^{\dag}\right)$. Here
$\gamma^\mu=(\gamma^0,\gamma^1)$ are the $\gamma$ matrices
satisfying the anticommutation relation
$\{\gamma^\mu,\gamma^\nu\}=2g^{\mu\nu}$. Since we have $1+1$
dimensions we are working with two-component spinors, and the
$\gamma$ matrices can be written in terms of the Pauli matrices.
Therefore, we choose the representation where $\gamma^0=-\sigma_1$
and $\gamma^1=i\sigma_2$. The mass matrix is

\begin{equation}
\label{eq:8} M=\frac{\varepsilon_{n}}{v_{F}^{2}}\gamma^0=
\left(\begin{array}{cc}0 &
-\frac{\varepsilon_{n}}{v_{F}^{2}}\\-\frac{\varepsilon_{n}}{v_{F}^{2}}
& 0\end{array}\right).
\end{equation}

The mass term in Eq.\ref{eq:7} indicates that a natural energy gap
given by the eigenvalues $(\Delta E = 2\varepsilon_{n})$ of the
mass-matrix appears in the system.

In order to make the model more realistic, the Peierls instability
is introduced in the two chains via spontaneous broken symmetry by
adding the term $|\phi|\overline{\Psi}\Psi$ to Eq.(\ref{eq:7}),
where $|\phi|$ is the constant value assumed by the scalar field out
of the kink-like defect. That scalar field represents the local
deformations of carbon ions induced by the Peierls instability. The
fermion Lagrangian is now

\begin{equation}
\label{eq:9} \mathcal{L}=i\hbar
v_F\overline\Psi\gamma^{\mu}\partial_{\mu}\Psi-v_{Fx}^2\overline\Psi
(M+|\phi|I)\Psi,
\end{equation}
where $I$ is the $2\times 2$ identity matrix. The new eigenvalues of the mass-matrix which appears in eq.(\ref{eq:9}) are $|\phi|\pm m_n$, where $m_n=\frac{\varepsilon_n}{v_{F}^2}$. The gap is shifted in the band structure in a way that the Fermi level is not divided symmetrically. But the value of the gap width remains the
same.

We can now define two chiral electronic eigenstates \cite{bernardini} as

\begin{equation}
|L\rangle\equiv\left(\begin{array}{r}1\\0\end{array}\right)_c\,\,\,\,\,{\text and}\,\,\,\,\,
|R\rangle\equiv\left(\begin{array}{r}0\\1\end{array}\right)_c.
\end{equation}
In this chiral representation the eigenstates are related to the electron itineracy to the
left and right, respectively. When normalized, these states are eigenstates of the
projection operator $P_{L,R}$. It is responsible for the projection of the pseudo-spins
along the electron propagation direction:

\begin{equation}
\label{eq:10} P_{L,R}=\pm\sigma_3.
\end{equation}
In the mass representation the mass operator $M+|\phi|I$, has eigenvectors

\begin{equation}
|{M}_1\rangle\equiv\left(\begin{array}{r}1\\0\end{array}\right)_m\,\,\,\,\,
{\text and}\,\,\,\,\,|{M}_2\rangle\equiv\left(\begin{array}{r}0\\1\end{array}\right)_m,
\end{equation}
that does not commute with $P_{L,R}$. As a result, chirality and
mass cannot be measured simultaneously. Therefore, we have two
representations for the electrons that can be related to each other
by the unitary transformation $U|M\rangle=|C\rangle$, where
$|C\rangle$ e $|M\rangle$ are general states of chirality and mass,
respectively. The simplest unitary matrix that can be used is

\begin{equation}
\label{eq:11}
U\equiv\left(\begin{array}{rr}\cos{\theta}&-\sin{\theta}\\\sin{\theta}&\cos{\theta}\end{array}\right),
\end{equation}
where we call $\theta$ a mixing angle.

Next, supposing the Hamiltonian is diagonalized in the mass representation

\begin{equation}
\label{eq:12}
\hat{H}_m=\left(\begin{array}{rr}E_1&0\\0&E_2\end{array}\right),
\end{equation}
where $E_1$ and $E_2$ are the possible electrons ``mass'' energies.
We can write the Hamiltonian in the chiral representation using the
similarity transformation $\hat{H}_c=U\hat{H}_m U^{\dag}$ or, in
terms of the Pauli matrices:

\begin{equation}
\label{eq:13}
\hat{H}_c=\frac{(E_2+E_1)}{2}I+\frac{(E_2-E_1)}{2}\Gamma
\end{equation}
where $\Gamma=\sigma_1\sin{2\theta}-\sigma_3\cos{2\theta}$, and $I$ is the
$2\times 2$ identity matrix. From the above equation we can see that $\hat{H}_c$ is not
diagonalized which implies that there is a transition between the chiral states left and
right. In the Schr\"odinger representation a chiral state can evolve in time from its
initial state by

\begin{equation}
\label{eq:14}
|C(t)\rangle=\exp{\left(-i\frac{\hat{H}_ct}{\hbar}\right)}|C\rangle,
\end{equation}
where $|C(0)\rangle=|C\rangle$, or writing in terms of the mixing angle $\theta$

\begin{eqnarray}
\label{eq:15}
|C(t)\rangle&=&\exp{\left[-i\frac{(E_1+E_2)t}{2\hbar}\right]}\times\\
&&(\cos \omega tI-i\sin \omega t\Gamma)|C\rangle\nonumber,
\end{eqnarray}
where $\omega=(\Delta E/2\hbar)$.

Considering that the electrons are propagating to the left in the initial time, {\it i.e.}
$|C\rangle=|L\rangle$, and knowing that

\begin{equation}
\label{eq:16}
\sigma_1|L\rangle=\left(\begin{array}{rr}0&1\\1&0\end{array}\right)
\left(\begin{array}{r}1\\0\end{array}\right)_c=\left(\begin{array}{r}0\\1\end{array}\right)_c
\end{equation}
and
\begin{equation}
\label{eq:17}
\sigma_3|L\rangle=\left(\begin{array}{rr}1&0\\0&-1\end{array}\right)
\left(\begin{array}{r}1\\0\end{array}\right)_c=\left(\begin{array}{r}1\\0\end{array}\right)_c
\end{equation}
we get the solution to the chiral state at any particular time

\begin{align}
\label{eq:18}
|C(t)\rangle&=\exp{\left[-i\frac{(E_1+E_2)t}{2\hbar}\right]}\left[(\cos \omega t+i\cos{2\theta}\sin \omega t)|L\rangle\right.\nonumber\\
&\left. -i\sin{2\theta}\sin \omega t|R\rangle\right].
\end{align}
The equation above shows that an electron initially in a ``left'' state
ends up in a superposition of ``left'' and ``right'' states. They oscillate and
after a time $t$ they have the probability of being in a chiral ``right" state given by

\begin{equation} \label{eq:19} P_{L\rightarrow R}=|\langle
R|C(t)\rangle|^2=\sin^2{2\theta}\sin^2\omega t,
\end{equation}
and the probability of them to return to the ``left" state is

\begin{equation}
\label{eq:20} P_{L\rightarrow L}=1-\sin^2{2\theta}\sin^2\omega t .
\end{equation}

We are now able to calculate the oscillation period of the
electrons. This period measures the time of one electron initially
in the left state goes to a right state and turn back to a left
state. That will happen when $t=\pi/\omega$. If we consider the
electron ``mass-energy" states with $(|\phi|\pm m_n)v_{Fx}^2$
eigenvalues, then $\Delta E=2m_n v_{Fx}^2$ (the gap width).
Therefore,

\begin{equation}
\label{eq:21} \mathcal{T}=\frac{\pi\hbar}{m_n
v_{Fx}^2}=\frac{\pi\hbar}{E}.
\end{equation}
The two possible directions of the electron propagation, its
chirality, are related to the two possible configurations
(enantiomers) that the molecule can have.

The energy needed for the electrons to overcome the distance $a$ between
the atoms is equal to the gap energy $(2m_n v_{Fx}^2 = 2E)$. Then, a
constant external force capable of making the oscillation must do a
work equal to $Fa$, it means $Fa = 2E$, or $E = Fa/2$. Substituting
that in Eq. 23 we have

\begin{equation}
\label{eq:22} \mathcal{T}=\frac{2\pi\hbar}{Fa}.
\end{equation}
This expression is exactly the same as the period of Bloch oscillations \cite{mossmann}.

In summary, we have shown that two paired and initially undimerized polyacetylene chains has an electronic energy band with a gap. Moreover, the pairing of the polyacetilene chains is also responsible for the quantum oscillation of the electron propagation. This oscillation comes from the fact that we cannot measure the chirality and mass simultaneously, and we have shown that it is equivalent to Bloch oscillations.

We thank M. G. Cottam and A. G. Souza Filho for useful discussions.
This work was supported by the Brazilian agencies CNPq and the
Funda\c c\~ao Cearense de Apoio \`a Pesquisa (FUNCAP).

\end{document}